# Large Grain Boundary Area Superconductors

G. Hammerl, H. Bielefeldt, S. Leitenmeier, A. Schmehl, C. W. Schneider, A. Weber, and J. Mannhart

*Experimentalphysik VI,*
*Center for Electronic Correlations and Magnetism,*
*Institute of Physics, Augsburg University, D-86135 Augsburg, Germany*

For many applications of polycrystalline high-$T_c$ superconductors the small critical currents of the grain boundaries pose a severe problem. To solve this problem we derive novel designs for the microstructure of coated conductors.

Most large scale applications of high-$T_c$ superconductors require the use of polycrystalline materials with large critical currents. In standard polycrystalline high-$T_c$ compounds the critical currents are limited by the grain boundaries [1]. To increase the critical currents, two strategies have been found. These strategies are based on enhancing the critical current density of the boundaries. The first consists in aligning the grains along all major axes to within few degrees [2]. This approach utilizes the fact that the grain boundary critical current density is a strongly decreasing function of the misorientation angle, dropping by three to four orders of magnitude as the misorientation angle is increased from 0° to 45° [1-4]. Second, for a given misorientation angle, the grain boundary critical current density is increased by appropriate doping [5,6]. To exploit the beneficial effects of grain alignment, the coated conductor technologies have been developed. Using ion beam assisted deposition (IBAD) [7,8], the rolling assisted biaxially textured substrate process (RABiTS) [9], or the inclined substrate deposition (ISD) [10,11], current densities above $10^6$ A/cm$^2$ have been achieved for conductors of several meter in length [12]. These current densities are further improved by the use of doping-multilayers [13].

To facilitate the complex fabrication of coated conductors, so that long tapes can be produced at competitive costs (see, e.g., [12]), many groups have devoted substantial efforts to enhance the critical currents of the coated conductors for a given grain alignment. Here we present a solution to this problem. This solution, derived from a straightforward analysis of grain boundary networks, is based on a novel design of the superconductor's microstructure.

The critical current of a grain boundary network does not equal the grain boundary critical current density multiplied by the geometrical cross-section $A$ of a superconductor because of two reasons [14,15]. First, current flow through the network is a complex, percolative process, possibly involving Josephson junctions and self field effects. Second, the effective area of the grain boundaries $A'$ may considerably exceed the cross-sectional area $A$. Therefore we envisaged that in coated conductors as well as in related polycrystalline superconductors, grain boundaries with large areas enable a significant enhancement of the critical currents. Such grain boundaries are formed, for example, by grains with fractal-like surfaces or by grains with large aspect ratios. In the latter case, which is the one considered in the following, the upper and lower sides as well as the lateral surfaces of the superconducting grains may be used to enhance the interface areas.

Elaborating these considerations, an astonishing effect is pointed out first:

Two coated conductors may be placed on top of each other such that the critical current of the stack exceeds the sum of the critical currents of the individual superconductors. Large critical currents are achieved by establishing an extended superconducting contact between the two superconducting surfaces.

This effect is obviously useful for the fabrication of coated conductors. Consider, for example, two such conductors, e.g. RABiTS- or IBAD-tapes, joined by welding the superconducting layers together as illustrated in Fig. 1. The resulting grain boundary areas are large compared to the interface areas available prior to the joining. The critical current enhancement is particularly notable for RABiTS-based superconductors, because in these tapes the grain lengths and widths equal tens of micrometers and therefore well exceed the typical film thickness of the order of one micrometer. Therefore large currents can pass from one superconducting layer to the other, even if the c-axis critical current densities are small and the grain boundary critical current densities are modest. Consequently, high critical current bypasses leading around the grain boundaries are created (see Fig. 1).

To further enhance the current densities, tapes covered on both sides with superconducting layers can be rolled, folded or stacked, examples of which are shown in Fig. 2.

To effectively utilize the lateral sides of the grains, we propose to grow the superconducting layers containing grains with large aspect ratios by using, for example, a RABiTS-type process. With standard metallurgical procedures such as rolling and subsequent annealing, metallic tapes with large aspect ratio grains are fabricated. These grains are then epitaxially replicated in the buffer layer system as well as in the superconducting layer (see Fig. 3). In such tapes, the supercurrent meanders within one superconducting layer as illustrated in Fig. 3. Again large critical currents are generated for given grain boundary critical current densities, because boundaries with large areas have been made available for the supercurrent flow. The RABiTS technology is particularly suitable for the implementations of grains with large aspect ratios. It is conceivable, however, that grains with large surface areas may also be produced using other techniques, such as IBAD or ISD.

To evaluate the influence of the aspect ratio of the grains on the critical current of such conductors, we have performed numerical calculations of the critical currents of model tapes consisting of $10^4$ to $10^6$ grains. These calculations were based on the mouse-algorithm developed by Holzapfel and coworkers [16]. In our calculations, for the intragrain critical current density a value of $5*10^6$ A/cm$^2$ was taken. The grain orientations were described by Gaussian distributions with full widths at half maximum (FWHM) of 4°, 8° and 25°, and the distributions were clipped at 45°. The grain lengths were chosen to follow clipped Gaussian distributions, too, with FWHM values of $l$/5 centered around the average length $l$. The results of the calculation, shown in Fig. 4, confirm that as anticipated the critical current of the superconductor increases monotonically with the grain aspect ratio, being limited ultimately by the intragrain critical current density only. Remarkably, the drop of the critical current density of the tape as a function of the average grain misorientation is significantly reduced for systems with large aspect ratios.

As shown by Fig. 4, the enhancement of the critical currents achievable by using grains with large surface areas are substantial. Of course, the maximal enhancements are obtained by combining a) grains with large aspect ratios to facilitate lateral transport and b) layered superconductors.

**Summary**
The percolation properties of grain boundary networks have been analyzed. Based on this straightforward analysis, designs of coated conductors have been proposed that use grains with large surface areas to enhance the critical currents. To attain competitive superconducting cables operating at 77K, we propose and urge to produce coated conductors that are based on a) grain alignment, b) selective doping, and c) grains with large aspect ratios.

Acknowledgement

The authors thank with pleasure J.G. Bednorz, P. Chaudhari and T. Claeson for useful discussions, as well as A. Herrnberger and Z. Ivanov for their support. Furthermore, we gratefully acknowledge interactions with U. Miller. This work was financially supported by the Bundesministerium für Forschung und Technologie (Project number 13N6918).

References

1  D. Dimos, P. Chaudhari, J. Mannhart, F. K. LeGoues, Phys. Rev. Lett. **61,** 219 (1988).

2  D. Dimos, P. Chaudhari, J. Mannhart, Phys. Rev. B **41,** 4038 (1990).

3  Z. G. Ivanov, P. Å. Nilsson, D. Winkler, J. A. Alarco, T. Claeson, E. A. Stepantsov, A. Ya. Tzalenchuk, Appl. Phys. Lett. **59,** 3030 (1991).

4  H. Hilgenkamp, J. Mannhart, Rev. Mod. Phys (in press).

5  A. Schmehl, B. Goetz, R. R. Schulz, C. W. Schneider, H. Bielefeldt, H. Hilgenkamp, J. Mannhart, Europhys. Lett. **47,** 110 (1999).

6  G. Hammerl, A. Schmehl, R. R. Schulz, B. Goetz, H. Bielefeldt, C. W. Schneider, H. Hilgenkamp, J. Mannhart, Nature **407,** 162 (2000).

7  Y. Iijima, N. Tanabe, O. Kohno, Y. Ikeno, Appl. Phys. Lett. **60,** 769 (1992).

8  X. D. Wu, S. R. Foltyn, P. N. Arendt, J. Townsend, C. Adams, I. H. Campbell, P. Tiwari, Y. Coulter, D. E. Peterson, Appl. Phys. Lett. **65,** 1961 (1994).

9  D. P. Norton, A. Goyal, J. D. Budai, D. K. Christen, D. M. Kroeger, E. D. Specht, Q. He, B. Saffian, M. Paranthaman, C. E. Klabunde, D. F. Lee, B. C. Sales, F. A. List, Science **274,** 755 (1996).

10 K. Hasegawa, N. Yoshida, K. Fujino, H. Mukai, K. Hayashi, K. Sato, T. Ohkuma, S. Honjyo, H. Ishii, T. Hara "In-plane aligned YBCO thin film tape fabricated by pulsed laser deposition," Proceedings of the International Cryogenic Engineering Conference (ICEC16), Kitakyushu, Japan (1996), p. 1413.

11 M. Bauer, R. Semerad, H. Kinder, IEEE Trans. Appl. Supercond. **9,** 1502 (1999).

12 R. F. Service, Science **295,** 787 (2002).

13 G. Hammerl, A. Weber, H. Bielefeldt, A. Schmehl, C. W. Schneider, J. Mannhart, 2002 (unpublished).

14 J. Mannhart, C. C. Tsuei, Z. Phys. B: Condens. Matter **77,** 53 (1989).

15 J. Mannhart "What limits the critical current density in high-$T_c$ superconductors?" in "Earlier and recent aspects of superconductivity" ed. by J. G. Bednorz and K. A. Müller (Springer-Verlag, Heidelberg), 208 (1990).

16 B. Holzapfel, L. Fernandez, F. Schindler, B. de Boer, N. Reger, J. Eickemeyer, P. Berberich, W. Prusseit, IEEE Trans. Appl. Supercond. **11,** 3872 (2001).

17 S. Leitenmeier, H. Bielefeldt, G. Hammerl, A. Schmehl, C. W. Schneider, J. Mannhart, 2002 (to be published in Annalen der Physik).

**Figure Legends**

**Fig. 1:** Example of a polycrystalline high-$T_c$ superconductor that exploits large grain boundary areas. Two RABiTS-tapes are joined such that an extended superconducting contact is established between the superconducting layers. This contact allows part of the supercurrent (arrow) to meander around grain boundaries.

**Fig. 2:** Two examples for implementations superconductors according to Fig. 1. Figure (a) shows a coated conductor with superconducting layers on both sides. Rolling of the tape produces an arrangement as shown in (a). The tape in (b) is a one-sided tape only, and the superconducting contact is established after folding the tape.

**Fig. 3:** Example of a coated conductor that uses large grain boundary areas by enlarging the side-areas of the grains. On top of a substrate tape which consists of grains with large aspect ratios, a buffer layer system and a superconductor are grown. The large grain boundary areas cause a large critical current for given critical current densities of the grain boundaries.

**Fig. 4:** Critical current density of a $YBa_2Cu_3O_{7-\delta}$-based coated conductor at 77 K as a function of the average grain aspect ratio, calculated for grain alignment spreads of 4°, 8° and 25°. Each datapoint reflects an average of at least 20 calculations. The black lines are guides to the eye (after [17]).

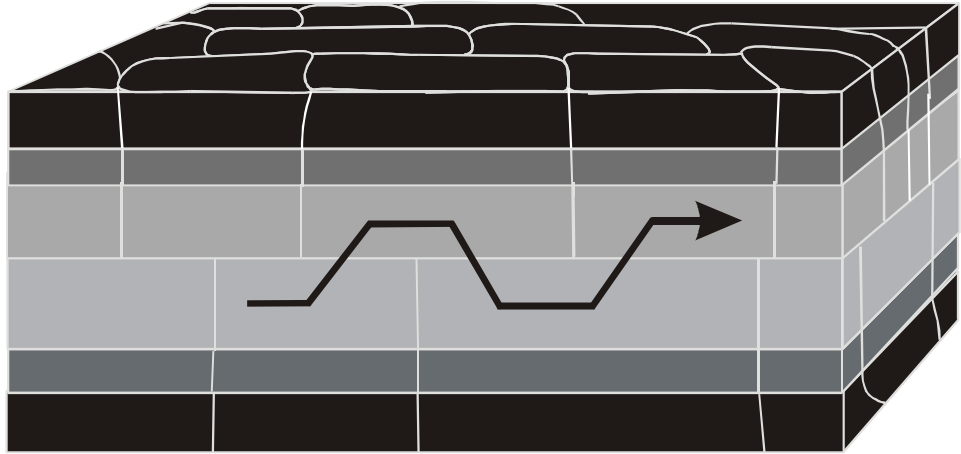

substrate tape    buffer layer    superconductor

a)

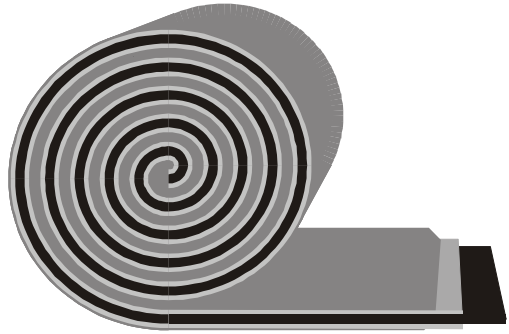

b)

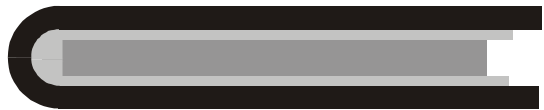

■ substrate tape     ■ buffer layer     ■ superconductor

*Hammerl et al., Fig. 2*

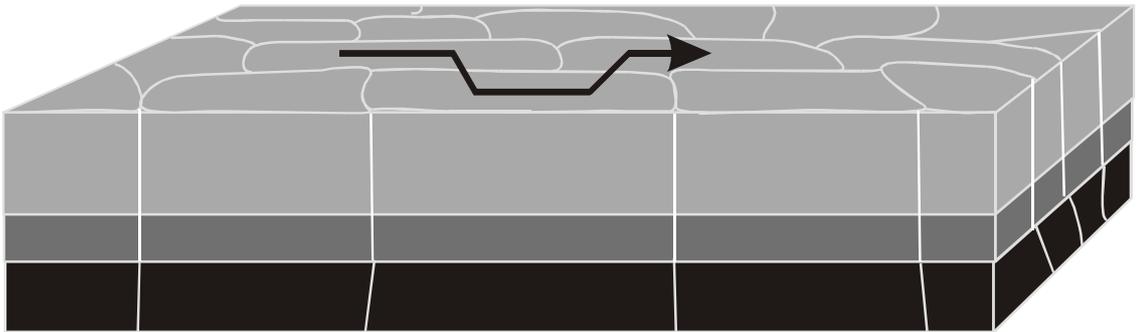

Hammerl et al., Fig. 3

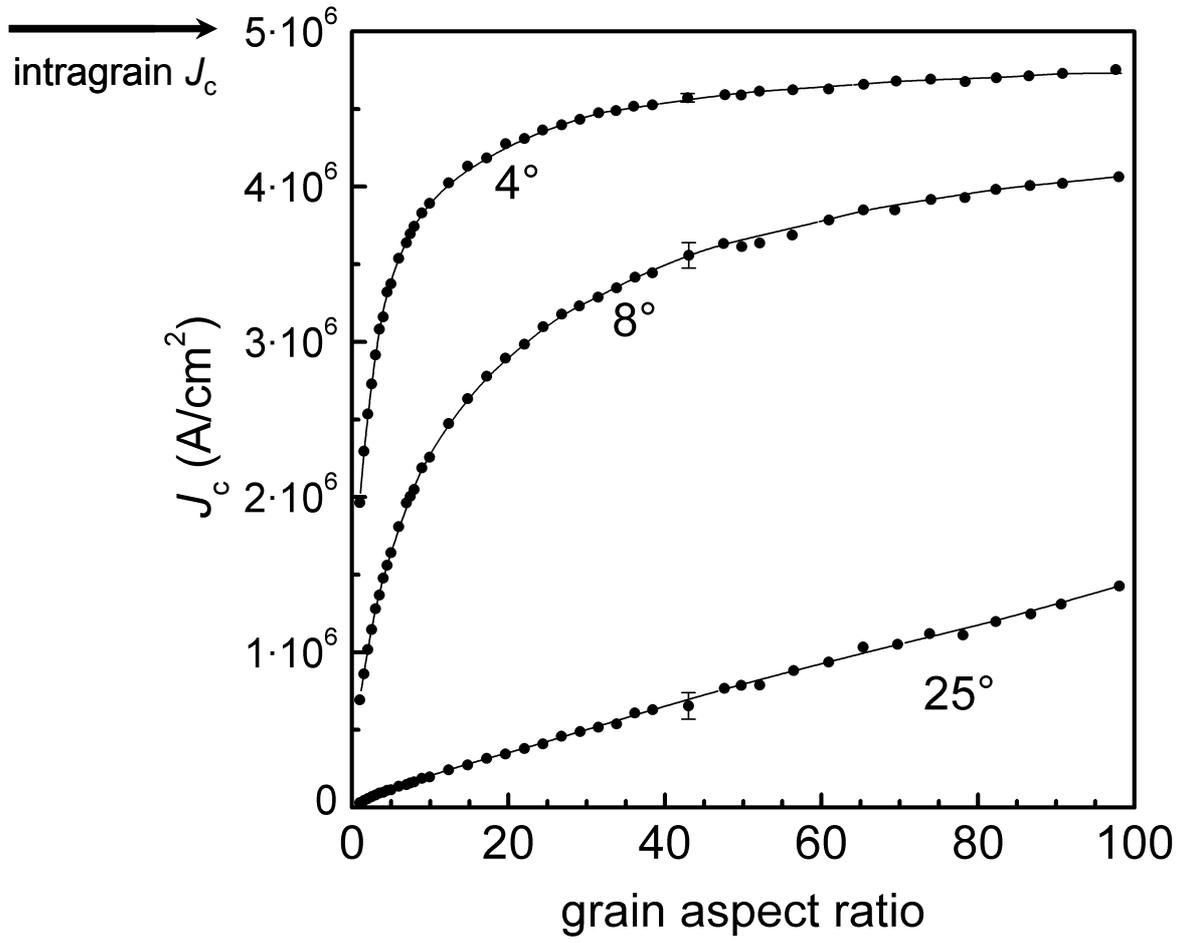

Hammerl et al., Fig. 4

11